\documentclass[preprint,amsmath,amssymb]{revtex4}

\usepackage{txfonts}    
\usepackage{graphicx}   
\usepackage{bm}

\newcommand{\sub}[1]{\hspace{-.07em}\bm{#1}}

\newcommand{\mr}[1]{\mathrm{#1} }

\newcommand{\vpr}{\varv_{\parallel} }

\newcommand{\vv}{\varv }

\newcommand{\dfg }{\delta f^{\mathrm{(g)}} }

\newcommand{\dphik }{\delta \phi_{\sub{k}_{\perp}} }

\begin{document}


\title{Isotope Effects on TEM-driven Turbulence and Zonal Flows in Helical and Tokamak Plasmas}

\author{Motoki Nakata$^\#$}
\author{Masanori Nunami}
\author{Hideo Sugama}
\affiliation{National Institute for Fusion Science, National Institutes of Natural Sciences, Toki 509-5292, Japan / \\
The Graduate University for Advanced Studies, Toki 509-5292, Japan}
\thanks{$\#\,$Present affiliation: Faculty of Arts and Sciences, Komazawa University, Tokyo 154-8525, Japan \\ 
\textbf{This is the author accepted manuscript (AAM/AM), published in Physical Review Letters 118, 165002 (2017)\\ doi: 10.1103/PhysRevLett.118.165002\ \ \ 
\copyright2017 American Physical Society }}
\author{Tomo-Hiko Watanabe}
\affiliation{Nagoya University, Nagoya 464-8602, Japan}

%

\begin{abstract}
Impacts of isotope ion mass on trapped electron mode (TEM) driven turbulence and zonal flows 
in magnetically confined fusion plasmas are investigated. 
Gyrokinetic simulations of TEM-driven turbulence in three-dimensional magnetic configuration of LHD plasmas 
with hydrogen isotope ions and real-mass kinetic electrons are realized for the first time, 
and the linear and the nonlinear nature of the isotope and collisional effects on the turbulent transport and zonal-flow generation is clarified. 
It is newly found that combined effects of the collisional TEM stabilization by the isotope ions 
and the associated increase in the impacts of the steady zonal flows at the near-marginal linear stability 
lead to the significant transport reduction with the opposite ion mass dependence in comparison to the conventional gyro-Bohm scaling.
The universal nature of the isotope effects on the TEM-driven turbulence and zonal flows is verified 
for a wide variety of toroidal plasmas, e.g., axisymmetric tokamak and non-axisymmetric helical/stellarator systems.  
\end{abstract}

\pacs{Valid PACS appear here}
\maketitle
\indent Turbulent transport and zonal-flow generation\cite{lin, fujisawa} are the universal physics 
in the drift-wave turbulence system in magnetized plasmas, 
and microinstabilities such as ion temperature gradient (ITG) modes 
and trapped electron modes (TEM) are typical causes of the turbulent transport to determine the confinement of energy, particle, 
and momentum in magnetically confined toroidal plasmas\cite{horton}. 
Impacts of the isotope ion mass in hydrogen (H), deuterium (D), and tritium (T) 
on the energy confinement\cite{BW_nf1993,cordey_nf1999,urano_prl,urano_nf2013,stroth}, and 
on the fluctuation characteristics\cite{xu,tj2} 
have been long-standing issues for several decades in plasma and fusion research, 
despite the broad interest and the importance for realizing sustainable fusion plasmas, which is one of the largest scientific challenges. \\
\indent A simple theoretical evaluation of the turbulent diffusivity $\chi_{\mr{turb.}}$ 
by the mixing-length diffusivity $\chi_{\mr{(ML)}}$ 
predicts a decrease of the energy confinement time $\tau_{E}$ ($\sim \! a^2/\chi_{\mr{turb.}}$)
due to the so-called gyro-Bohm scaling indicating the isotope ion mass dependence of $\sqrt{A_{\mr{i}}}$ 
in the mixing-length diffusivity, i.e., 
$\chi_{\mr{(ML)}} \! \sim \! (\gamma/k_{\perp}^2) \! \propto \! (\sqrt{A_{\mr{i}}}/Z_{\mr{i}}^2)\chi_{\mr{GB(H)}}$ 
($\gamma$: the growth rate of microinstability, $k_{\perp}$: the wavenumber, $A_{\mr{i}}$: the mass number of ion species i, 
$Z_{\mr{i}}$: the charge number, $\chi_{\mr{GB(H)}}$: gyro-Bohm diffusivity for hydrogen ion). 
However, many experimental studies in tokamak devices such as ASDEX\cite{BW_nf1993}, JET\cite{cordey_nf1999}, 
and JT-60U\cite{urano_prl,urano_nf2013} have shown, more or less, 
the opposite isotope mass dependence leading to improved energy confinement, e.g., 
$\tau_{E} \! \propto \! A_{\mr{i}}^{0.5}$, depending on the heating and profile conditions. 
In stellarator devices, a slightly moderate dependence of $\tau_{E} \! \propto \! A_{\mr{i}}^{0.2 \pm 0.15}$ has been observed 
in high-$T_{\mr{e}}/T_{\mr{i}}$ discharges with the electron-cyclotron resonance heating (ECRH)\cite{stroth}. 
Also, some influences on the long-range turbulence correlation in the toroidal direction, 
which is regarded as a proxy of the zonal flow, have been observed\cite{xu,tj2}. 
The first-principle-based multi-species gyrokinetic simulations with isotope ions 
are powerful approaches to establishing theoretical understanding of the underlying physical mechanisms, 
as well as further investigations 
of the isotope effects by forthcoming hydrogen/deuterium experiments in, e.g., Large Helical Device (LHD), 
Wendelstein 7-X (W7-X), JT-60SA, and ITER. \\
\indent Many efforts have been devoted so far to theoretical and numerical studies on the ITG-driven turbulent transport 
and the zonal-flow generation, based on the gyrokinetic model.  
In particular, several recent works have addressed the isotope effects on the ITG-dominated turbulence 
in tokamak plasmas\cite{pusztai_pop2011, bustos_pop2015}. 
However, the simulation results indicate a slight deviation from the gyro-Bohm mass dependence of the turbulent transport level,  
depending on the impurity species and the profile gradient regime, and did not reproduce the experimental tendency 
of $\chi_{\mr{turb.}} \! \propto \! A_{\mr{i}}^{\alpha},\ \alpha \! < \! 0$. 
On the other hand, for the non-axisymmetric plasmas such as LHD and W7-X, 
it has been found that the three-dimensional nature of the magnetic field leads 
to the isotope and/or global effects on the turbulent transport and the zonal-flow generation 
through the coupling with the equilibrium radial electric field\cite{sugama_pop2009er,watanabe_nf2011,Mishchenko,pavlos}. 
The roles of the equilibrium radial electric field in the enhancement of zonal flows 
have also been investigated in several stellarator experiments\cite{pedrosa, wilcox}. \\
\indent In addition to the ITG modes, the TEM is often destabilized in a wide variety of toroidal plasmas\cite{helander} with high density gradients 
and/or electron temperature gradients, and the TEM-driven turbulent transport is another important transport channel in burning plasmas 
with the strong electron heating by the $\alpha$ particles.   
More recent gyrokinetic validation studies have identified a TEM-dominated state in the outer-core region 
of the L-mode plasmas\cite{nakata_nf2016, told}. 
Also, the impact of the TEM on the pedestal-top region has been found in the global simulation\cite{Fulton}. 
The fundamental properties of the collisionless TEM instability and the turbulent transport, including the impact of the zonal flow, 
have been investigated by analytic\cite{proll} and numerical\cite{merz,lang,proll_opt} studies. 
As for the isotope effects on the TEM, although there are several linear analyses based on fluid\cite{tokar1} 
and kinetic\cite{nakata_ppcf2016,idomura_jcp2016} models, the nonlinear behavior, including the zonal-flow generation,  
has not been fully understood yet. \\
\indent In this study, the isotope ion mass impacts on the TEM-driven turbulence and zonal flows 
in helical and tokamak plasmas are explored by using gyrokinetic simulations 
with real-mass kinetic electrons and the finite collisions.   
It is, for the first time, elucidated that the linear TEM stabilization by the isotope ions and the associated increase 
of the zonal-flow impact lead to the transport reduction distinct from the conventional gyro-Bohm mass dependence. \\
%
%
\indent The linear and nonlinear analyses are carried out by an electromagnetic gyrokinetic Vlasov code GKV\cite{gkv,watanabe_prl}
with multiple-particle-species treatment\cite{nakata_cpc2015} and realistic MHD equilibria in tokamak and helical plasmas. 
The time evolution of perturbed gyrocenter distribution functions $\dfg_{\mr{s}}$ (``s'' denotes the particle species) 
in the five-dimensional phase space ($x, y, z, \vpr, \mu$) is numerically solved 
under a fixed equilibrium with the local Maxwellian $F_{\mr{Ms}}$, 
where the so-called fluxtube coordinates of $x \! = \! a(\rho - \rho_0)$, 
$y \! =\! (a\rho_0/q(\rho_0))[q(\rho)\theta - \zeta]$, and 
$z \! =\! \theta$ are used ($q$: the safety factor, $a$: the plasma minor radius). 
Both the electrostatic ($\dphik$) and electromagnetic ($\delta A_{||k}$) fluctuations 
in the wavenumber space with $\bm{k}_{\perp} = k_{x}\nabla x + k_y \nabla y$ 
are treated in GKV simulations, but we focus on the low-$\beta$ plasmas, so that the magnetic fluctuation is quite small. 
Collisional effects are introduced in terms of an Lenard-Bernstein type linearized collision operator $\mathcal{C}_{\mr{s}}$. 
The normalized collision frequency is defined as  
$\nu_{\mr{ss^{\prime}}}^{\ast} \! = \! qR_{\rm{ax}}\tau_{\mr{ss^{\prime}}}^{-1}/(\!\sqrt{2}\epsilon^{3/2}\vv_{\mr{ts}})$ 
with the characteristic collision time $\tau_{\mr{ss^{\prime}}}$, the inverse aspect ratio $\epsilon$, 
and the thermal speed $\vv_{\mr{ts}} \! =\! (T_{\mr{s}}/m_{\mr{s}})^{1/2}$. 
The scale lengths of the logarithmic gradients for the density and temperature are given by $R_{\mr{ax}}/L_{n_{\mr{s}}}$ 
and $R_{\mr{ax}}/L_{T_{\mr{s}}}$, respectively. 
More detailed descriptions for the simulation model and the technical terms are given 
in, e.g., Refs. \cite{nakata_ppcf2016} and \cite{nakata_pfr2014}. 
%
%
\begin{figure}
\includegraphics[scale=0.9]{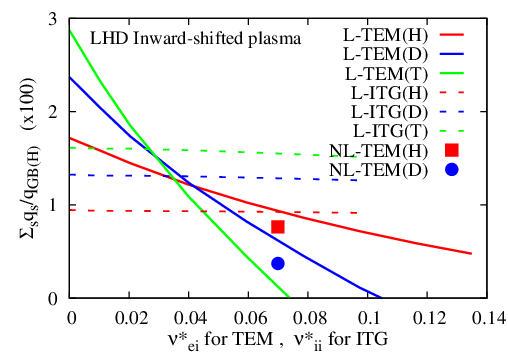}
\caption{Collisionality dependence of the radial heat flux $\sum_{\mr{s=i,e}}q_{\mr{s}}$ 
with the mixing-length diffusivity $\gamma/k_{\perp}^2$ for the linear TEM (L-TEM) and ITG (L-ITG) modes 
in LHD inward-shifted H-, D-, and T-plasmas, where $k_{x}\rho_{\mr{ts}} \! = \! 0$, $k_{y}\rho_{\mr{ts}} \! = \! 0.4$ (s = \{H, D, T\}). 
For qualitative comparison, the nonlinear results from Fig. 2(a) are also displayed for the TEM case (NL-TEM). 
}
\end{figure}
%
It should be stressed that a good prediction capability for the ion and electron energy fluxes has been confirmed 
in the validation study against the actual JT-60U tokamak\cite{nakata_nf2016} and LHD experiments\cite{nunami_pop2013}. \\
%
\begin{figure}
\includegraphics[scale=0.9]{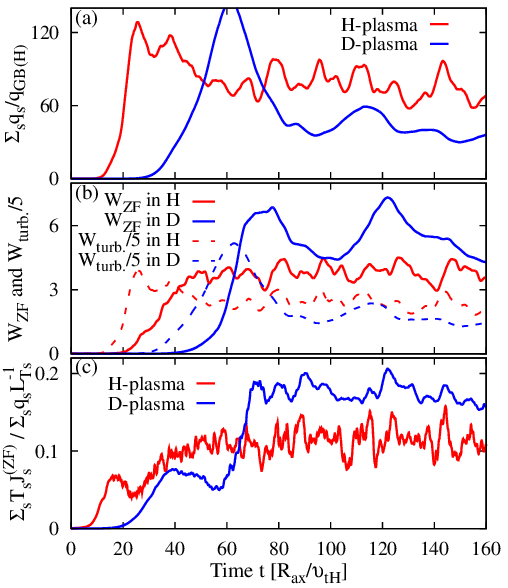}
\caption{
Impacts of the hydrogen isotope mass on (a) turbulent heat fluxes normalized by the hydrogen gyro-Bohm unit ($\sum_{\mr{s}}q_{\mr{s}}/q_{\mr{GB(H)}}$), (b) turbulence ($W_{\mr{turb.}}$) and zonal-flow ($W_{\mr{ZF}}$) energy, and (c) normalized entropy transfer 
to zonal flows ($\sum_{\mr{s}} T_{\mr{s}}\mathcal{T}^{\mr{(ZF)}}_{\mr{s}}/\sum_{\mr{s}}q_{\mr{s}}L_{T_{\mr{s}}}^{-1}$), 
where s = \{i, e\}, and $\nu^{\ast}_{\mr{ei}} \! = \! 0.07$.      
}
\end{figure}
%
\indent In order to estimate the isotope ion mass impacts on the turbulent transport levels from the linear calculations, 
the collisionality dependencies of the radial heat flux 
$\sum_{\mr{s}=\mr{i,e}}q_{\mr{s}} \! = \! -n_{\mr{s}}a\chi_{\mr{s}}g^{\rho \rho}dT_{\mr{s}}/d\rho$ 
for hydrogen- (H-), deuterium- (D-), and tritium- (T-) plasmas are shown in Fig. 1, 
where $g^{\rho \rho}$ is the geometric factor, and the heat diffusivity 
$\chi_{\mr{s}}$ is evaluated by the mixing-length diffusivity $\gamma/k_{\perp}^2$ 
for the fastest growing mode ($k_{\mr{x}}\rho_{\mr{ts}}\! =\! 0,\ k_{y}\rho_{\mr{ts}}\! =\! 0.4$) in the linear calculations, 
and the hydrogen gyro-Bohm diffusivity $\chi_{\mr{GB(H)}} \! = \! \rho_{\mr{tH}}^2\vv_{\mr{tH}}/R_{\mr{ax}}$ 
with the thermal gyroradius $\rho_{\mr{ts}} = m_{\mr{s}}\vv_{\mr{ts}}/eB_{\mr{ax}}$ is used as the unit. 
Here, we focus on the inward-shifted case in the LHD plasma with $R_{\mr{ax}}\! =\! 3.60\mr{m}$, 
$q(\rho_0) \! = \! 1.70$, $\hat{s}(\rho_0) \! = \! -0.96$, 
and $a\rho_0 /R_{\mr{ax}} \! = \! 0.082$, which shows the higher TEM growth rate compared with that in the standard configuration.       
The equilibrium profile parameters used here are
\{$R_{\mr{ax}}/L_{T_{\mr{i}}}\! =\! 10$, $R_{\mr{ax}}/L_{T_{\mr{e}}} \! =\! 12$, $R_{\mr{ax}}/L_{n}\! =\! 2$, $T_{\mr{e}}/T_{\mr{i}}\! =\! 1$\} 
and  
\{$R_{\mr{ax}}/L_{T_{\mr{i}}}\! =\! 1$, $R_{\mr{ax}}/L_{T_{\mr{e}}} \! =\! 12$, $R_{\mr{ax}}/L_{n}\! =\! 2$, $T_{\mr{e}}/T_{\mr{i}}\! =\! 2.5$\} 
for ITG- and TEM-dominated cases, respectively, 
where these parameters are based on the high-temperature discharges in the actual LHD experiments, 
and the detailed linear spectra of the growth rate and mode frequency are given in Ref. \cite{nakata_ppcf2016}.  
For the cases in ITG modes (labeled by L-ITG in the figure) showing weak $\nu^{\ast}_{\mr{ii}}$ dependence, 
we see that the linear results with the mixing-length diffusivity exhibit a gyro-Bohm ion mass dependence, i.e., 
$\gamma/k_{\perp}^{2} \! \propto \! \sqrt{A_{\mr{i}}}$. 
The similar $\sqrt{A_{\mr{i}}}$ dependence is also found in the collisionless limit of TEM (labeled by L-TEM). 
This means that the TEM growth rate in the collisionless limit is scaled with the ion transit frequency of 
$\omega_{\mr{ti}} \! = \! \vv_{\mr{ti}}/q_{0}R_{\mr{ax}}$.  
However, a different reduction tendency of the TEM cases depending on the isotope ion species is revealed 
in the finite collisionality regime, 
where the ion mass dependence in the ratio of the electron-ion collision frequency to the ion transit frequency, 
i.e., $\nu_{\mr{ei}}/\omega_{\mr{ti}} \! \propto \! (m_{\mr{i}}/m_{\mr{e}})^{1/2} \! \propto \! \sqrt{A_{\mr{i}}}$, 
leads to stronger collisional stabilization for heavier isotope ions. 
Note also that the stabilization effect on ITG modes by the ion-ion collisions is almost independent of the ion mass, 
i.e., $\nu_{\mr{ii}}/\omega_{{\mr{ti}}} \! \propto m_{\mr{i}}^{0}$. 
Then, the opposite ion mass dependence of $A_{\mr{i}}^{\alpha}$ with $\alpha \! < \! 0$ appears for the TEM 
in a certain collisionality regime, i.e., $\nu^{\ast}_{\mr{ei}} \! \geqslant \! 0.04$ in the present case. 
The reduction in the mixing-length diffusivity for the TEM through the collisional effects 
provides us with a useful qualitative basis to investigate the isotope impacts on the turbulent transport. 
For the comparison, which will be shown below, the nonlinear TEM results (labeled by NL-TEM) are also plotted in Fig. 1. \\
%
%
\indent As for the linear zonal-flow response, earlier theoretical works show the weak isotope effects, 
i.e., the zonal-flow response kernel indicates no explicit mass dependence for the fixed $k_{\perp}\rho_{\mr{ti}}$\cite{sugama_prl2005}, 
except for the cases with the equilibrium radial electric field\cite{sugama_pop2009er,watanabe_nf2011,Mishchenko}. 
%
%
\begin{figure}
\includegraphics[scale=0.35,clip]{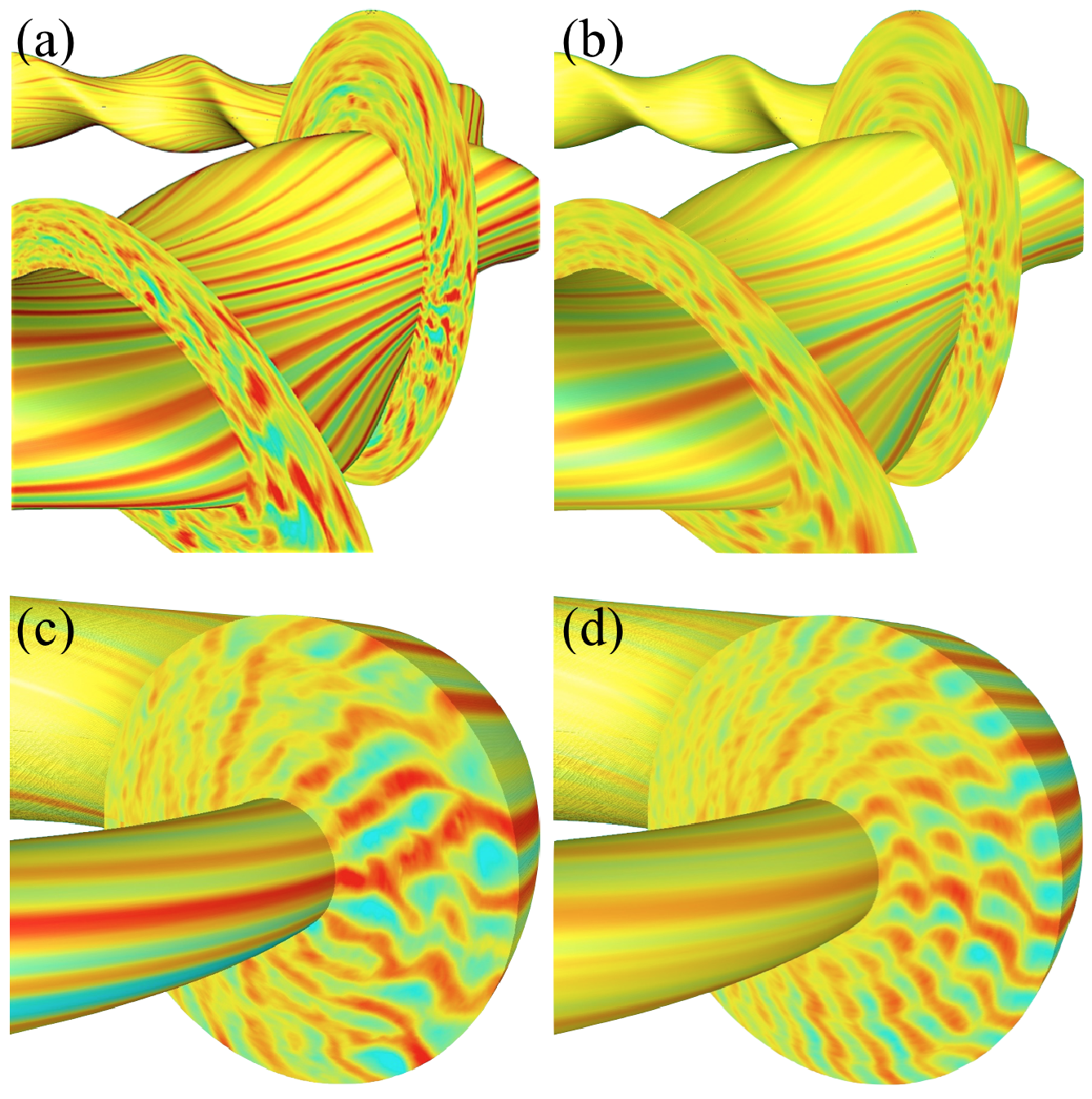}
\caption{Spatial structures of the potential fluctuations in the TEM-driven turbulence for (a) LHD H-, (b) LHD D-, 
(c) CBC-like tokamak H-, and (d) CBC-like tokamak D-plasmas, where the simulation conditions of LHD and CBC tokamak 
cases correspond to those in Figs. 2 and 4(b) ($\nu^{\ast}_{\mr{ei}}\! =\! 0.035$), respectively. }
\end{figure}
%
Therefore, the isotope mass impacts on the nonlinear zonal-flow dynamics become more important. \\
%
\indent Following the above linear analyses, we have performed massively parallel nonlinear TEM-driven turbulence simulations 
for the non-axisymmetric LHD plasma, where $\sim \! 200$ hours with 69,120 computation cores are required for a hydrogen case.   
The equilibrium parameter set for TEM shown above is used, and $\nu^{\ast}_{\mr{ei}} \! = \! 0.07$ is considered. 
We employ a number of grid points in ($x, y, z, \vpr, \mu$) as ($256 \times 96 \times 320 \times 90 \times 24$) 
for the ion and electron, 
where sufficiently large box sizes of $L_{x} \! = \! 147.3\rho_{\mr{tH}}$ (or $k_{x\mr{(min)}}\rho_{\mr{tH}} \! = \! 0.04265$) 
and $L_{y} = 148.1\rho_{\mr{tH}}$ (or $k_{y\mr{(min)}}\rho_{\mr{tH}} \! = \! 0.04243$) are taken 
for the comparison between H- and D-plasmas. \\
\indent Nonlinear GKV simulation results on the time evolution of the turbulent radial heat flux $\sum_{\mr{s}=\mr{i,e}}q_\mr{s}$, 
the turbulence energy $W_{\mr{turb.}}$, and the zonal-flow energy $W_{\mr{ZF}}$
are shown in Figs. 2(a) and 2(b), 
where $W_\mr{turb.}$ and $W_\mr{ZF}$ are defined as the non-zonal ($k_y\! \neq\! 0$) and the zonal ($k_y\! = \! 0$) components of 
$W_{\mr{total}} \! = \! \langle \sum_{k_x,k_y}(e_{\mr{s}}^2n_{\mr{s}}/2T_{\mr{s}})[1-\Gamma_{0}(k_{\perp}^2\rho_{\mr{ts}}^2)]
|\dphik|^2  \rangle_z 
\! \simeq \! \langle \sum_{k_x,k_y}(n_{\mr{i}}T_{\mr{i}}/2)k_{\perp}^{2}\rho_{\mr{ti}}^2
(e^2|\dphik|^2/T_{\mr{i}}^2) \rangle_z$ with the field-line-averaging operator $\langle \cdots \rangle_z$. 
Also, the time evolution of the normalized entropy transfer from turbulence to zonal modes, 
$\sum_{\mr{s}=\mr{i,e}}T_{\mr{s}}\mathcal{T}_{\mr{s}}^{\mr{(ZF)}}/\sum_{\mr{s}=\mr{i,e}}q_{\mr{s}}L_{T_{\mr{s}}}^{-1}$, 
is shown in Fig. 2(c), 
where $\mathcal{T}_{\mr{s}}^{\mr{(ZF)}}$ is regarded as a kinetic extension of the zonal-flow energy production 
due to the Reynolds stress (see Ref. \cite{nakata_pop2012} for the definitions), 
and should balance with the collisional dissipation for the zonal modes, $\mathcal{D}_{\mr{s}}^{\mr{(ZF)}}$, 
in the statistically steady turbulence state. 
%
%
\begin{figure}
\includegraphics[scale=0.9]{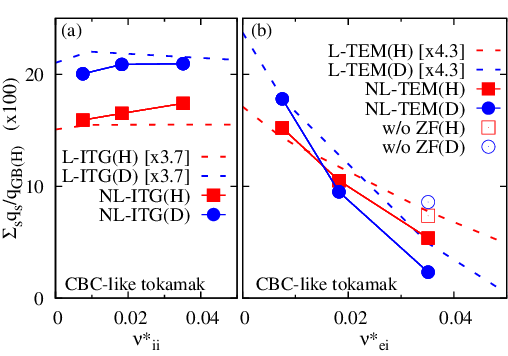}
\caption{Collisionality dependence of the radial heat flux $\sum_{\mr{s}}q_{\mr{s}}$ for (a) ITG and (b) TEM cases in CBC-like tokamak plasmas, 
where the linear mixing-length (L-) and the corresponding nonlinear (NL-) results are plotted by dashed-lines and line-symbols, respectively. 
The linear results are uniformly scaled for qualitative comparison of profiles. NL-TEM cases without the zonal flow are also displayed 
by the open symbols.}
\end{figure}
%
Actually, the time averaged entropy balance relation of 
$\sum_{\mr{s}=\mr{i,e}}T_{\mr{s}}\overline{\mathcal{T}_{\mr{s}}^{\mr{(ZF)}}} \! 
+ \! \sum_{\mr{s}=\mr{i,e}}T_{\mr{s}}\overline{\mathcal{D}_{\mr{s}}^{\mr{(ZF)}}}=0$ (the overline means the time average) 
is accurately satisfied within a relative error less than 10\% in the present TEM turbulence simulations. 
As shown in Fig. 2(a) and also by symbols in Fig. 1, the turbulent transport level in the D-plasma is lower than that in the H-plasma, 
where the ratio of the mean turbulent transport level is evaluated as 
[$\sum_{\mr{s}}q_\mr{s}$ for D]/[$\sum_{\mr{s}}q_\mr{s}$ for H] = 0.48, which exhibits more significant reduction than that 
in the linear estimation with the ratio of 0.66. 
One also finds that the zonal-flow energy $W_{\mr{ZF}}$ increases in the D-plasma 
in spite of the slight decrease of the turbulence energy $W_{\mr{turb.}}$, 
where the zonal-flow enhancement characterized by the increase of the mean relative amplitude\cite{watanabe_nf2011, nunami_pop2013}, 
$\overline{W_{\mr{ZF}}/W_{\mr{total}}}$, from 0.25 in H-plasma to 0.39 in D-plasma is well correlated 
with the normalized entropy transfer shown in Fig. 2(c). 
The effective zonal-flow shearing rate defined by $\omega_{\mr{ZF}}/\gamma_{\mr{max}}$ in the D-plasma is more than twice 
as large as that in the H-plasma, 
where $\omega_{\mr{ZF}} = \langle |\partial_{x}\vv_{\mr{ZF}}| \rangle_{x}$ means the shearing rate of the steady zonal flows, 
and $\langle \cdots \rangle_{x}$ denotes the spatial average in the radial ($x$) direction. 
As will be shown below, the increase of the zonal-flow impact at the near-marginal stability in the D-plasma is induced 
by the stronger collisional stabilization of the TEM for heavier isotope ions. \\
\indent Three dimensional spatial structures of the potential fluctuations are shown in Figs. 3(a) -- 3(d), 
where hydrogen and deuterium cases are displayed. 
For comparison, the results of Cyclone-base-case (CBC) like tokamak TEM cases (shown below) are also presented. 
%
%
\begin{figure}
\includegraphics[scale=0.9]{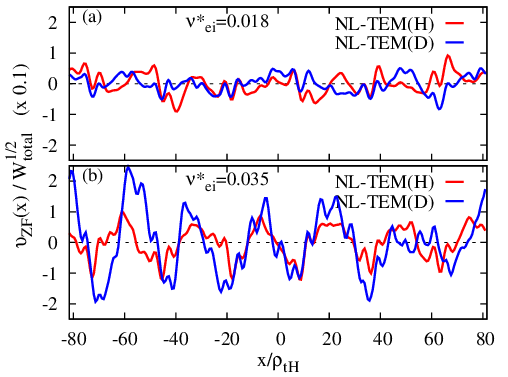}
\caption{Radial profiles of TEM-driven zonal flows in CBC-like tokamak H- and D-plasmas 
for (a) $\nu^{\ast}_{\mr{ei}} \! = \! 0.018$ and (b) $\nu^{\ast}_{\mr{ei}} \! = \! 0.035$, 
where a larger relative amplitude of the steady zonal flow appears in D-plasma with $\nu^{\ast}_{\mr{ei}} \! = \! 0.035$ in comparison to the others.  
}
\end{figure}
%
We see more decorrelated structures in the radial direction with lower fluctuation amplitudes for both the helical and tokamak D-plasmas 
with relatively stronger zonal-flow generation. \\
%
%
\indent It should be stressed that since the collisional stabilization of the TEM shown in the present analyses 
is not particular for helical plasmas, 
the similar isotope effects on the turbulent transport and zonal flows are expected in tokamak plasmas, as well. 
For verifying this, the nonlinear simulations for CBC-like tokamak plasmas are performed, 
where the equilibrium profile parameters of 
\{$R_{\mr{ax}}/L_{T_{\mr{i}}}\! =\! 8$, $R_{\mr{ax}}/L_{T_{\mr{e}}} \! =\! 8$, $R_{\mr{ax}}/L_{n}\! =\! 3$, $T_{\mr{e}}/T_{\mr{i}}\! =\! 1$\} 
and 
\{$R_{\mr{ax}}/L_{T_{\mr{i}}}\! =\! 1$, $R_{\mr{ax}}/L_{T_{\mr{e}}} \! =\! 8$, $R_{\mr{ax}}/L_{n}\! =\! 3$, $T_{\mr{e}}/T_{\mr{i}}\! =\! 3$\} 
are used for the ITG- and the TEM-dominated cases, respectively. 
The so-called $s$ -- $\alpha$ toroidal geometry with $q(\rho_0) \! = \! 1.42$, $\hat{s}(\rho_0) \! = \! 0.8$, and $a\rho_0 /R_{\mr{ax}} \! = \! 0.18$ 
is considered. 
The GKV simulation results are summarized in Figs. 4(a) and 4(b), which are similar as the Fig. 1.  
As is discussed in the above LHD case, the similar collisionality dependence
is found for the linear tokamak ITG cases [labeled by L-ITG in Fig. 4(a)] with the gyro-Bohm mass dependence, 
and for the linear tokamak TEM cases [labeled by L-TEM in Fig. 4(b)] with the opposite dependence on the isotope ion mass 
around $\nu^{\ast}_{\mr{ei}} \! = \! 0.025$. 
It is also revealed that the $\nu^{\ast}_{\mr{ii}}$-dependence of turbulent heat flux in the nonlinear ITG cases (labeled by NL-ITG) 
is qualitatively similar to that in the linear cases with the $\sqrt{A_{\mr{i}}}$-dependence, 
where $\overline{W_{\mr{ZF}}/W_{\mr{total}}} \! \sim \! 0.2$ for all of the ITG cases. 
In contrast, as $\nu^{\ast}_{\mr{ei}}$ increases, the nonlinear TEM results (labeled by NL-TEM) show steeper decrease of the transport level 
in comparison to the linear cases, 
where [$\sum_{\mr{s}}q_\mr{s}$ for D]/[$\sum_{\mr{s}}q_\mr{s}$ for H] = 0.43 in the nonlinear TEM case for $\nu^{\ast}_{\mr{ei}} \! = \! 0.035$ 
indicating more significant reduction than that in the linear estimation with the ratio of 0.63. \\ 
\indent The transport reduction beyond the linear TEM stabilization is also attributed to the nonlinear turbulence suppression 
by the zonal flows as the TEM growth rate decreases towards the marginal stability with increasing $\nu^{\ast}_{\mr{ei}}$, 
as is similar to those in the near-marginal collisionless (or weakly collisional) ITG turbulence\cite{rogers, nakata_nf2013}. 
Indeed, Figs. 5(a) and 5(b) show the radial profiles of the long-time averaged steady zonal flows in the TEM turbulence for H- and D-plasmas, 
where the cases with and without the transport reduction by the isotope effects 
(or correspondingly, $\nu^{\ast}_{\mr{ei}} \! = \! 0.018$ and $\nu^{\ast}_{\mr{ei}} \! = \! 0.035$) are compared. 
It is found that the relative amplitude of the steady zonal flows is enhanced in the near-marginal TEM stability[Fig. 5(b)], 
while the lower and comparable amplitudes are found in H- and D-plasmas with relatively higher TEM growth rates [Fig. 5(a)].  
Also, the effective zonal-flow shearing $\omega_{\mr{ZF}}/\gamma_{\mr{max}}$ 
significantly increases from 0.76 in H-plasma to 2.90 in D-plasma 
with $\nu^{\ast}_{\mr{ei}} \! = \! 0.035$, whereas it shows only slight increase from 0.44 to 0.63 for the cases with $\nu^{\ast}_{\mr{ei}} \! = \! 0.018$. 
Although the relative amplitude of the TEM-driven zonal flows, e.g., $\overline{W_{\mr{ZF}}/W_{\mr{total}}} \! = \! 0.048$ in D-plasma 
with $\nu^{\ast}_{\mr{ei}} \! = \! 0.035$, is much lower than that in the ITG cases, 
the numerical suppression of the zonal-flow components in the nonlinear simulation leads to more than 4 times the enhancement of the transport level 
[shown by the open symbols in Fig. 4(b)], which implies the crucial impact of the zonal flows in the near-marginal TEM stability.     
In the other cases with larger TEM growth rates, weaker zonal-flow impacts, which are pointed out in Ref. \cite{lang} for the collisionless TEM cases, have also been confirmed.  \\
%
%
%
\indent In summary, this Letter presents a critical mechanism of the isotope effects on the TEM instability, 
based on the systematic linear gyrokinetic analyses with finite collisions for both tokamak and helical plasmas. 
The ion mass dependence appearing in the ratio of electron-ion collision frequency to the ion transit frequency, 
which gives the characteristic frequency of the collisionless TEM, leads to
the reduction in the TEM growth rate for the case with heavier isotope ions.  
The nonlinear gyrokinetic simulations have, for the first time, identified 
the significant transport reduction with the opposite ion mass dependence in comparison to the conventional gyro-Bohm scaling, 
through the isotope effects of the collisional TEM stabilization and the steady zonal flows. 
It is noteworthy that the present isotope effects on the TEM are expected in a wide variety of toroidal plasmas 
including tokamak and helical/stellarator plasmas, particularly for the outer-core and/or pedestal-top region with large density and/or 
electron-temperature gradients.  
In non-axisymmetric plasmas the equilibrium radial electric field, which is ignored here, 
gives rise to an additional isotope effect on the zonal-flow enhancement\cite{pedrosa,wilcox,sugama_pop2009er,watanabe_nf2011,Mishchenko}. 
In order to verify the present isotope mass effects, we need a precise setup of 
the fluctuation measurements and profile control to complement the previous experimental results. 
It should be stressed, though, that the reduction of the TEM-driven turbulent transport predicted here is qualitatively consistent 
with the experimental observations in ASDEX L-mode plasmas with $\tau_{E} \! \propto \! A_{\mr{i}}^{0.5}$, 
where the increase of $\tau_{E}$ by the isotope effect becomes more pronounced 
in a higher electron density regime\cite{BW_nf1993}. 
The forthcoming hydrogen/deuterium experiments in stellarator and tokamak devices, such as LHD, W7-X, and JT-60SA, 
are capable of investigating the isotope mass effects quantitatively, 
and our gyrokinetic analyses provide a possible operation scenario with the improved confinement. \\  
\indent This work is supported by the MEXT Japan, Grant Nos. 26820401, 26820398, and 16K06941, in part by the NIFS collaborative Research Programs 
(NIFS16KNST096, NIFS16KNTT036, NIFS16KNST095, NIFS16KNST099, and NIFS16KNTT035), and in part by the MEXT grant for Post-K project: 
Development of innovative clean energy, Core design of fusion reactor. 
Numerical simulations were performed by Plasma Simulator at NIFS, HELIOS at IFERC-CSC, and K-computer at RIKEN-AICS 
on the HPCI System Research project (hp160117).  
%
%
%
%


\end{document}